\documentclass[usenatbib]{mnras}
\usepackage{amssymb,amsmath}
\usepackage{graphicx,booktabs}
\usepackage[T1]{fontenc}
\usepackage{ae,aecompl}
\usepackage[normalem]{ulem}
\newcommand{\be}{\begin{equation}}
\newcommand{\ee}{\end{equation}}

\title[]{Grammage of cosmic rays in the proximity of  supernova remnants embedded in a partially ionized medium}

\author[S. Recchia et al.]
{S. Recchia$^{1,2}$\thanks{E-mail: sarah.recchia@unito.it}, 
D. Galli$^3$,
L. Nava$^4$,
M. Padovani$^3$,
S. Gabici$^5$,
A. Marcowith$^6$,
\newauthor
V. Ptuskin$^7$,
G. Morlino$^3$
\\
$^1$Dipartimento di Fisica, Universit\'a di Torino, via P. Giuria 1, 10125 Torino, Italy\\
$^2$Istituto Nazionale di Fisica Nucleare, Sezione di Torino, Via P. Giuria 1, 10125 Torino, Italy\\
$^3$INAF–Osservatorio Astrofisico di Arcetri, Largo E. Fermi 5, 50125 Firenze, Italy\\
$^4$INAF-Osservatorio Astronomico di Brera, Via Bianchi 46, I-23807 Merate, Italy\\
$^5$Universit\'e de Paris, CNRS, Astroparticule et Cosmologie, F-75006 Paris, France\\
$^{6}$Laboratoire Univers et particules de Montpellier, Universit\'e Montpellier/CNRS, F-34095 Montpellier, France\\
$^{7}$Pushkov Institute of Terrestrial Magnetism, Ionosphere and Radiowave Propagation, 108840, Troitsk, Moscow, Russia
}

\date{Accepted XXX. Received YYY; in original form ZZZ}

\begin{document}
%\label{firstpage}
\maketitle

\begin{abstract}
We investigate the damping of  Alfv\'en waves generated by the
cosmic ray resonant streaming instability in the context of the 
cosmic ray escape and propagation in the proximity of supernova 
remnants. We consider ion-neutral
damping, turbulent damping and non linear Landau damping in the
warm ionized and warm neutral phases of the interstellar medium.
For the ion-neutral damping, up-to-date damping coefficients are used.
We investigate in particular whether the self-confinement of cosmic 
rays nearby sources can appreciably affect the grammage. 
We show that the ion-neutral
damping and the turbulent damping effectively limit the residence
time of cosmic rays in the source proximity, so that the grammage
accumulated near sources is found to be negligible. Contrary to previous results, this also
happens in the most extreme scenario where ion-neutral damping
is less effective, namely in a medium with  only neutral helium
and fully ionized hydrogen.
Therefore, 
the standard picture, in which CR secondaries are produced
during the whole time spent by cosmic rays throughout the Galactic
disk, need not to be deeply revisited.
\end{abstract}

\begin{keywords}
%hydrodynamics -- ISM: clouds -- ISM: kinematics and dynamics
\end{keywords}

%--------------------------------------------------
\section{Introduction}
\label{sec:intro}
%-------------------------------------------------

The most popular hypothesis for the origin of Galactic cosmic rays
(CRs) invokes supernova remnants (SNRs) as the main sources of such
particles (see e.g. \citealt{Blasi-2013-review,
Gabici-2019-review-con-altri}). In this scenario, which in the last
decades had become a paradigm, CR diffusion plays a central role.
Diffusion is the key ingredient at the base of the  diffusive shock
acceleration of particles at SNRs \citep[e.g.][]{Drury-1983}.
Diffusion also affects the escape of CRs from the acceleration site and
the subsequent propagation in the source region, with prominent
implications for $\gamma$-ray observations
(\citealt{Aharonian-Atoyan-1996-gamma-SNR, Gabici-2009-gamma-MC,
Casanova-2010, Ohira-2011-gamma-MC, Nava-2013-anisotropic}). Finally,
diffusion determines the confinement time of CRs in the Galaxy,
thus affecting the observed spectrum and the abundances of secondary
spallation nuclei and of unstable isotopes
\citep{PtuskinSoutoul1998,Wiedenbeck2007}.

The diffusion of CRs is thought to be mostly due to the resonant
scattering off plasma waves 
whose wavelength is comparable to the particle's
Larmor radius $r_{\rm L}
=\gamma m_p c^2/e B$, where $m_p$ is the proton mass, $B$ is the magnetic field strength and $\gamma$ the Lorentz factor
(see e.g. \citealt{Skilling-1975-I}).
The  magneto-hydrodynamic (MHD)   turbulence relevant for CR propagation is composed of incompressible Alfv\'enic and compressible (fast and slow) magnetosonic fluctuations \citep{Cho-Lazarian-2002, Fornieri-Gaggero-2021}. MHD turbulence is ubiquitous in the interstellar space and may be injected by astrophysical sources (see e.g. \cite{MacLow-2004}) but also by CRs themselves.
The active role of CRs in producing the waves responsible for their
scattering has been widely recognized (see e.g. \citealt{Wentzel1974,
Skilling-1975-III, Cesarsky1980, Amato2011}). In fact, spatial
gradients in the CR density, as those found in the source vicinity,
lead to the excitation of Alfv\'en waves at the resonant scale
\citep{Ptuskin-2008}. This process, called \textit{resonant streaming
instability}, produces waves that propagate along magnetic
field lines in the direction of decreasing CR density.

The density of Alfv\'en waves that scatter CRs is limited by several
damping processes. The most relevant are: 
({\it i}\/) ion-neutral
damping in a partially ionized medium (\citealt{Kulsrud-1969,Kulsrud-1971, Zweibel-1982-ion-neutral});
({\it ii}\/) turbulent damping, due to the interaction of a
wave with counter-propagating Alfv\'en wave packets. Such waves
may be the result of a background turbulence injected on large
scales and cascading to the small scales (we indicate this damping
as FG, after \citealt{Farmer-2004}); 
({\it iii}\/) non-linear
Landau (NLL) damping, due to the interaction of background
thermal ions with the beat of two interfering Alfv\'en waves (see
e.g. \citealt{Felice2001, Wiener-2013-WIM-ion-neutral}).
The relative importance of these effects depends significantly on
the physical conditions and chemical composition of the ambient medium.
A few other collisionless and collisional damping processes
can impact magnetohydrodynamical wave propagation in a partially
ionized gas but they mostly affect high-wavenumber perturbations
\citep{Yan-Lazarian-2004-compressible}. Recently, it has  been suggested that dust grains may also contribute to the damping of Alfv\'en waves \citep{Squire-2021MNRAS.502.2630S}.

In this paper we investigate the escape of CRs from SNRs, and their
subsequent self confinement in the source region, as due to the
interplay between the generation of Alfv\'en waves by CR streaming
instability, and the damping process mentioned above. 
Our main goal  is to establish whether the
self-confinement of CRs nearby sources can appreciably affect the
grammage accumulated by these particles. In fact, if this is the
case, a significant fraction of CR secondaries would be produced
in the vicinity of CR sources, and not during the time spent
by CRs in the Galactic disk, as commonly assumed. This would
constitute a profound modification of the standard view of CR
transport in the Galaxy \citep[see, e.g.][]{Dangelo-2016-grammage}. 
In particular, we focus 
on the CR propagation in partially ionized phases of the 
interstellar medium (ISM), showing that the ion-neutral
and FG damping can significantly affect the residence time 
of CRs nearby their sources. We find that, for typical conditions,
the grammage accumulated by
CRs in the vicinity of sources is negligible compared to that
accumulated during the time spent in the Galaxy. 
Even in the case of a medium made 
of fully ionized H and neutral He, the combination of ion-neutral and turbulent damping can substantially 
affect the confinement 
time\footnote{The case of a
fully neutral (atomic or partially molecular) medium and of a diffuse molecular medium \citep[see, e.g.][]{Brahimi2020} 
are not treated here, since the filling factor of
such phases is small, but we report the ion-neutral
damping rate for such media for the sake of completeness. The
case of a fully ionized medium has been extensively treated by \cite{Nava-2019-esc-II}.}.

This paper is organized as follows: in Sec.~\ref{sec:damp} we 
describe the damping of Alfv\'en waves by ion-neutral collisions
in various partially ionized phases of the ISM, and by other damping mechanisms; in Sec.~\ref{sec:cr-prop}
we illustrate the equations and the setup of our model of CR escape
and propagation in the proximity of SNRs, 
the time dependent CR spectrum and diffusion coefficient, the
residence time of CRs in the source proximity and the implications
on the grammage; 
in Sec.~\ref{sec:res} we describe our results;
and finally in Sec.~\ref{sec:conc} we draw our conclusions.

\section{Damping of Alfv\'en waves}
\label{sec:damp}

\subsection{Ion-neutral damping}
\label{dampingaw}

The Galaxy is composed, for most of its volume, by three ISM phases,
namely the warm neutral medium (WNM, filling factor $\sim 25\%$),
warm ionized medium (WIM, filling factor $\sim 25$\%) and hot
ionized medium (HIM, filling factor $\sim 50$\%, see e.g.
\citealt{Ferriere-2001, Ferriere-2019}). The physical characteristics of these
phases are summarised in
Table~\ref{table:ISM} (from \citealt{jean2009}, see also
\citealt{Ferriere-2001, Ferriere-2019}). 
The physical characteristics of the cold neutral
medium (CNM) and the diffuse medium (DiM) are also listed for
completeness, while their filling factor is $\lesssim 1\%$ \citep{Ferriere-2001, Ferriere-2019}.  In the regions where neutrals are present, like the
WNM and the WIM, the rate of ion-neutral damping depends on the
amount and chemical species of the colliding particles.  In the WNM
and WIM the ions are H$^+$, while neutrals are He atoms (with a
H/He ratio of $\sim~10$\%) and H atoms with a fraction that varies
from phase to phase.

\begin{table*}
\centering
\caption{ISM phases and parameters adopted in this work.
$T$ is the gas temperature, $B$ the interstellar magnetic field, $n$ the total gas density, $f$ the ionisation fraction, $\chi$ the helium fraction and $L_{\rm inj}$ the injection scale of the background magnetic turbulence.}
\resizebox{\linewidth}{!}{
\begin{tabular}{lrccccccc}
\toprule\toprule
       & $T$\,(K) & $B$\,($\mu$G) & $n~(\mbox{cm}^{-3})$ & neutral & ion & $f$ & $\chi$ & $L_{\rm inj}$ (pc) \\
\midrule
WIM  &  8000  & 5  & 0.35   & H, He  & H$^+$  & 0.6$-$0.9        & 0$-$0.1 \\
         &            &      &          & He       & H$^+$  & 1                 & 0.1 & 50 \\   
\midrule
WNM &  8000  & 5  & 0.35  & H, He  & H$^+$  & $7\times10^{-3}-5\times10^{-2}$ & 0$-$0.1 & 50  \\         
\midrule
CNM  &  80      & 5  & 35     & H, He & C$^+$  &  $4\times 10^{-4} - 10^{-3}$                  &  0.1 & 1-50 \\
\midrule
DiM  &  50  & 5  & 300   & $\rm H_2$, He  & C$^+$  &   $10^{-4}$         & 0.1 & 1-50  \\
\midrule
HIM  & $10^6$  & 5  & $\sim 0.01$   & - & H$^+$  &  1.0         & 0.0 & 100  \\
\bottomrule
\end{tabular}
}
\label{table:ISM}
\end{table*}

The main processes of momentum transfer (mt) between ions and neutrals
are elastic scattering by induced dipole, and charge 
exchange (ce). In
the former case, dominant at low collision energies, the incoming
ion is deflected by the dipole electric field induced in the neutral
species, according to its polarizability (Langevin scattering); in the latter case the incoming ion takes one or more electrons from
the neutral species, which becomes an ambient ion.  The friction
force per unit volume ${\bf F}_i$ exerted on an ion $i$ is thus the
sum of ${\bf F}_{i,{\rm mt}}+{\bf F}_{i,{\rm ce}}$.

With the exception of collisions between an 
ion and a neutral of the same species, as 
in the important case of collisions of H$^+$ ions with H atoms (see Sec.~\ref{subsec:coll-HH-HHe}), the two processes are well separated in energy. At low collision energies elastic scattering dominates, and the friction 
force is 
\be
{\bf F}_{i,{\rm mt}}=n_i n_n \mu_{in}\langle\sigma_{\rm mt} v\rangle_{in} ({\bf u}_n-{\bf u}_i),
\label{fricel}
\ee
where $n_i$ and $n_n$ are the ion and neutral densities, ${\bf u}_i$ and ${\bf u}_n$ 
are the ion and neutral velocities, $\mu_{in}$ is the reduced mass of the colliding
particles, $\sigma_{\rm mt}$ is the momentum transfer (hereafter m.t.) cross section,
and the brackets denote an average over the relative velocity of the colliding particles.\\

At high collision energies (above $\sim 10^2$~eV), the dominant contribution to the 
transfer of momentum is charge exchange
\be
{\rm A}^+ + {\rm B} \rightarrow {\rm A} + {\rm B}^+.
\ee
If the charge exchange rate coefficient is approximately independent of temperature,
and there is no net backward-forward asymmetry in the scattering process (two conditions
generally well satisfied), \cite{Draine-1986} has shown that the friction force on the 
ions takes the form
\be
{\bf F}_{i,{\rm ce}}=n_i n_n \langle\sigma_{\rm ce} v\rangle_{in} \frac{m_n^2{\bf u}_n-m_i^2{\bf u}_i}{m_n+m_i},
\label{fricce}
\ee
where $\sigma_{\rm ce}$ is the charge exchange (hereafter c.e.)
cross section, and $m_{n(i)}$ the mass of the neutral (ion).

The collisional rate coefficients $\langle\sigma_{\rm mt} v\rangle_{in}$ 
and $\langle\sigma_{\rm ce} v\rangle_{in}$ 
are often 
estimated from the values
given by \cite{Kulsrud-1971} or \cite{Zweibel-1982-ion-neutral} for H$^+$-- H collisions 
\citep[e.g.][]{Dangelo-2016-grammage, Nava-2016-esc-I, Brahimi2020}.
The rate coefficients for collisions between various species of ions and neutrals adopted 
in this study are described in detail in Sec.~\ref{subsec:coll-HH-HHe}. 
For elastic collisions, 
they have been taken from the 
compilation by \cite{pg08}; for charge 
exchange, they have been calculated from the most updated available cross sections.

Ion-neutral collisions are one of the dominant
damping processes for Alfv\'en waves propagating in a partially ionized medium
(see \citealt{p56, Kulsrud-1969}). In the case of elastic ion-neutral collisions, (Eq.~\ref{fricel}),  the dispersion relation for Alfv\'en waves in this case 
is
\be
\omega(\omega^2-\omega_k^2)+i \nu_{in}[(1+\epsilon)\omega^2-\epsilon\omega_k^2]=0,
\label{disp}
\ee
where $\omega$ is the frequency of the wave, $\omega_k=k v_{{\rm A},i}$ 
is the wavevector in units of the Alfv\'en speed of the ions
\be
v_{{\rm A},i}=\frac{B}{\sqrt{4\pi m_i n_i}}, 
\ee
$\nu_{in}$ 
is the ion-neutral collision frequency
\be
\nu_{in}= \frac{m_n}{m_i+m_n}\langle\sigma_{\rm mt} v\rangle_{in} n_n,
\label{nusum}
\ee
and $\epsilon$ is the ion-to-neutral mass density ratio
\be
\epsilon=\frac{m_i n_i}{m_n n_n}.
\ee
Notice that $\epsilon$ is a small quantity in the WNM and CNM but not in the WIM\footnote{To be precise,
the dispersion relation Eq.~(\ref{disp}) is valid only if
the friction force is proportional to the ion-neutral relative speed ${\bf
u}_n-{\bf u}_i$, as in the case of momentum transfer by elastic collisions. However,
we use the same relation also in the case of charge exchange, simply replacing 
$\langle\sigma_{\rm mt} v\rangle_{in}$ 
with 
$\langle\sigma_{\rm ce} v\rangle_{in}$.}.

The dispersion relation Eq.~(\ref{disp}) is a cubic equation for the wave frequency $\omega$ (with real and imaginary parts) as a function of the real wavenumber $\omega_k$. Writing $\omega=\Re(\omega)-i\Gamma_d^{in}$, where $\Gamma_d^{in} >0$ is the ion-neutral 
damping rate, and substituting in Eq.~(\ref{disp}), one obtains 
\citep{Zweibel-1982-ion-neutral}
\be
\omega_k^2=\frac{2\Gamma_d^{in}}{\nu_{in} - 2\Gamma_d^{in}}[(1+\epsilon)\nu_{in}-2\Gamma_d^{in}]^2,
\ee
which implies $0<\Gamma_d^{in} < \nu_{in}/2$.
%If $\Gamma_d^{in} \ll \nu_{in}/2$, then
If $\epsilon \ll 1$, then
\be
\Gamma_d^{in}\approx\frac{\omega_k^2\nu_{in}}{2[\omega_k^2+(1+\epsilon)^2\nu_{in}^2]}.
\ee
Alfv\'en waves resonantly excited by CR protons have frequency
$\omega_k\approx v_{{\rm A},i}/r_{\rm L}$
Thus, the frequency is related to the kinetic energy 
of the CR proton $E=\gamma m_p c^2$ as  
\be
\omega_k\approx \frac{eB v_{{\rm A},i}}{E}.
\ee
The effective Alfv\'en velocity, $v_{\rm A} = \Re(\omega)/k$, felt by CRs depends 
on the coupling between ions and neutrals. In general, the  following asymptotic behavior can be identified:
\begin{itemize}
\item Low wavenumber, $\omega_k \ll \nu_{in}$:\\ at large CR energy
ions and neutrals are well coupled; 
the total density is $n=n_{\rm H}+n_{\rm He}+n_i$ and
    the Alfv\'en
    speed relevant for CRs resonant with the waves is 
    \be
    v_{{\rm A},n}=\frac{B}{\sqrt{4\pi \mu m_p n}},
    \label{vatot}
    \ee
    where $\mu \sim 1.4$ is the mean molecular weight,  and $\Gamma_d^{in} \propto E^{-2}$;
\item High wavenumber, $\omega_k \gg \nu_{in}$:\\ at small CR energy
    ions and neutrals are weakly coupled and ion-neutral damping is most
    effective. The   Alfv\'en speed is the one in the
    ions, $v_{{\rm A},i}$,
    and $\Gamma_d^{in}
    \sim \rm const$.
\end{itemize}

Notice that if $\epsilon < 1/8$ there is a range of wavenumbers for
which the waves do not propagate in a partially ionized medium
\citep{Zweibel-1982-ion-neutral}. This is marked as a shaded region
in Fig.~\ref{fig:damping_WNM_WIM}-\ref{fig:damping_CNM}.
On the
other hand, such non-propagation band is found in the absence of CRs propagating in the partially ionized medium. Recently it has been suggested \citep{Reville-2021-damping}
that taking into account the presence of CRs may  allow for the propagation of waves in that band.

Introducing the fraction of ionized gas
$f$ and the helium-to-hydrogen ratio $\chi$,
\be
f=\frac{n_i}{n_{\rm H}+n_i}, \qquad \chi=\frac{n_{\rm He}}{n_{\rm H}+n_i},
\ee
Eq.~(\ref{nusum}) becomes
\be
\nu_{in}=
\left[\frac{1-f}{1+\tilde{m}_i}\langle\sigma_{\rm mt} v\rangle_{i,{\rm H}}
+\frac{4\chi}{4+\tilde{m}_i}\langle\sigma_{\rm mt} v\rangle_{i,{\rm He}}\right]
\frac{n}{1+\chi}.
\ee
where $\tilde{m}_i=m_i/m_p$. In the following, the standard value 
$\chi=0.1$ is assumed, but the case $\chi=0$ is also considered for illustrative 
purposes and for a comparison with the results of~\cite{Dangelo-2016-grammage}, who neglect the contribution of helium to ion-neutral damping.

\subsubsection{WIM and WNM}

\begin{figure*}
\includegraphics[width=\columnwidth]{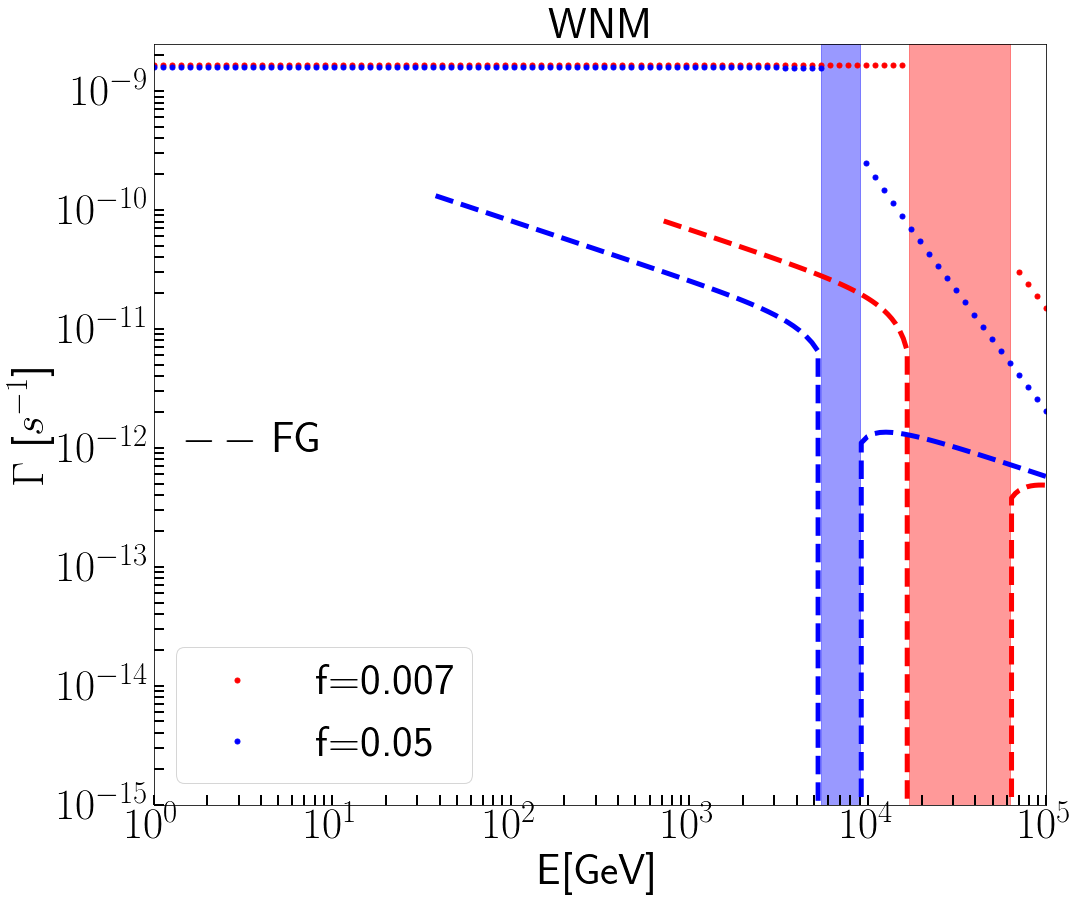}
\includegraphics[width=\columnwidth]{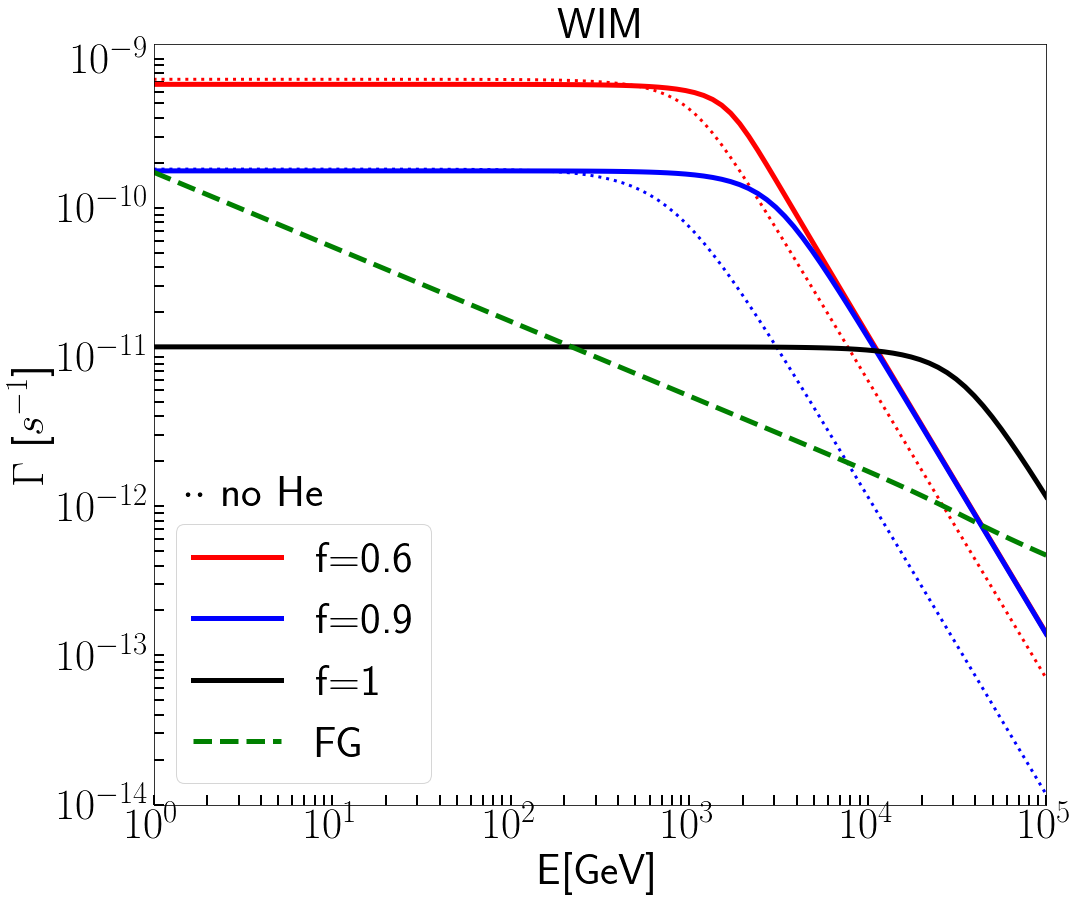}
\caption{Damping rates $\Gamma_d^{in}$ and $\Gamma_d^{\rm FG}$ (ion-neutral and turbulent) of Alfv\'en waves in the
WNM ({\it left-hand panel}\/) and WIM ({\it right-hand panel}\/) vs. CR energy $E$. Different colors are used for different values of the hydrogen ionization fraction $f$. Unless stated otherwise, a standard 
$\chi = 0.1$ He abundance is assumed.
The considered parameters for the WNM and WIM are given in
Table~\ref{table:ISM}. 
In the left-hand panel the dotted lines refers to ion-neutral damping, while the dashed lines to the FG damping. The last are truncated to the minimum energy, $E_{\rm min}$,  below which the background turbulence  is damped by ion-neutral friction before reaching the scale relevant for damping self-generated waves resonant with particle of energy $< E_{\rm min}$ (see Sec.~\protect\ref{subsec:FG}).
The shaded regions represent the
range of non-propagation of Alfv\'en waves (see Sec.~\protect\ref{dampingaw}).
In the right-hand  panel, the solid lines refer to ion-neutral damping, while the  thin dotted lines refer to the case  where the contribution to damping from He is neglected
($\chi =0$). The dashed lines refer to the FG damping, which in the case of WIM is found to depend little on the ionization fraction $f$. 
}
\label{fig:damping_WNM_WIM}
\end{figure*}

In this case H is partially ionized and the dominant ion is H$^+$ ($\tilde{m}_i=1$). 
Therefore $\epsilon=n_{{\rm H}^+}/
(n_{\rm H}+4n_{\rm He})=f/(1-f+4\chi)$, i.e. $\epsilon=0.005$--$0.05$ and 0.75--9 
for the WNM and the WIM, respectively. The ion-neutral collision frequency is
\be
\nu_{in}=\left[\frac{1-f}{2}
\langle\sigma_{\rm mt} v\rangle_{{\rm H}^+,{\rm H}} 
+\frac{4\chi}{5}\langle\sigma_{\rm mt} v\rangle_{{\rm H}^+,{\rm He}}\right] \frac{n}{1+\chi}.
\ee
Fig.~\ref{fig:damping_WNM_WIM} shows the damping rate for waves resonant with 
CRs of energy $E$, as a function of the CR energy $E$.
Notice the non-propagation band found in the WNM ($\epsilon < 1/8$).

\subsubsection{CNM and DiM}

\begin{figure*}
% \centering {
 \includegraphics[width=\columnwidth]{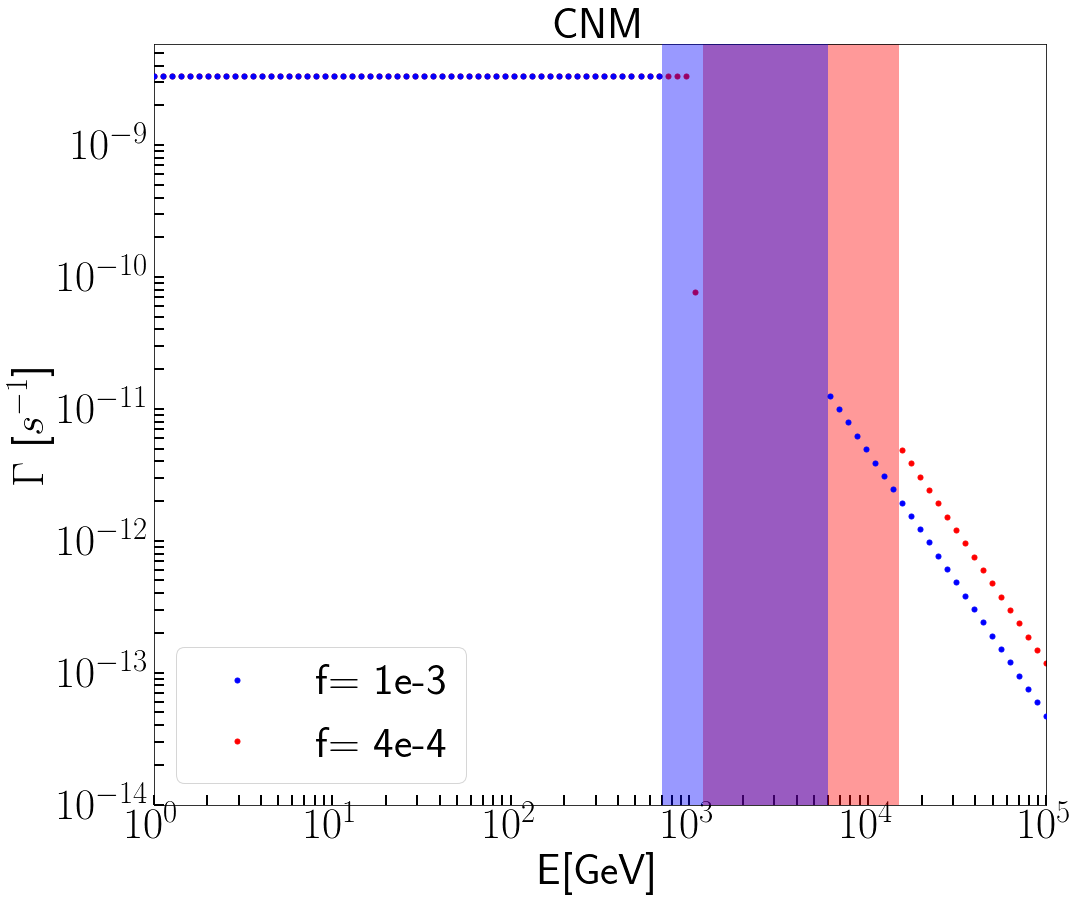}
 \includegraphics[width=\columnwidth]{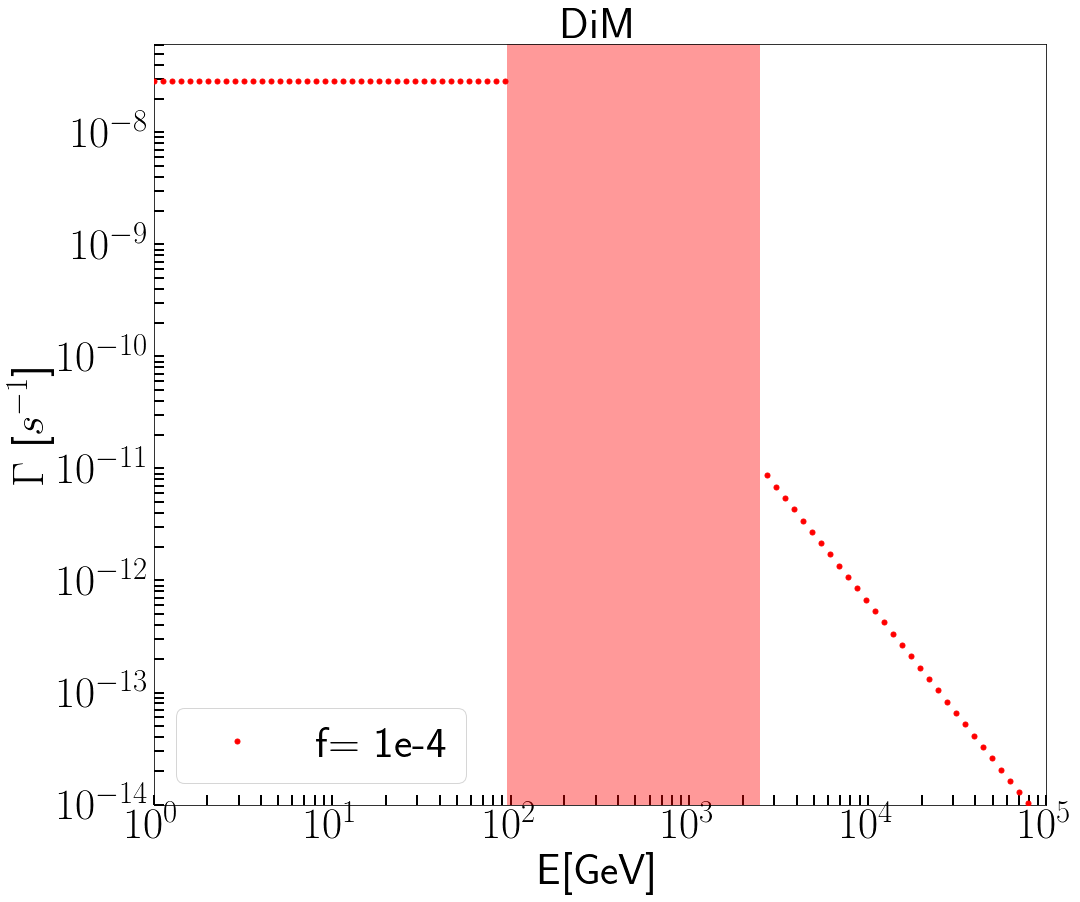}
 %}
\caption{Ion-neutral damping rate $\Gamma_d^{in}$ of Alfv\'en waves in the CNM
(left-hand panel) and DiM (right-hand panel) as a function of the
CR energy $E$. Different colors are used for different values of
the C$^+$ abundance. The parameters
adopted for the CNM and DiM are given in Table~\ref{table:ISM}. The
shaded regions represent the range of non-propagation of Alfv\'en
waves (see Sec.~\ref{dampingaw}).}
\label{fig:damping_CNM}
\end{figure*}

In this case H is neutral and the dominant ion is C$^+$ ($\tilde{m}_i=12$), 
with fractional abundance 
$n_{{\rm C}^+}/n_{\rm H}\approx (0.4$--$1)\times 10^{-3}$. 
Therefore $\epsilon=12 n_{{\rm C}^+}/(n_{\rm H}+4n_{\rm He})\approx (3$--$9)\times 10^{-3}$ and
\be
\nu_{in}=\left[\frac{1}{13}\langle\sigma_{\rm mt} v\rangle_{{\rm C}^+,{\rm H}}
+\frac{\chi}{4}\langle\sigma_{\rm mt} v\rangle_{{\rm C}^+,{\rm He}}\right] \frac{n}{1+\chi}.
\ee
Fig.~\ref{fig:damping_CNM} shows the damping rate for waves resonant with CRs of 
energy $E$, as a function of the CR energy $E$. Also in this case non-propagation regions are found.

\subsection{Wave cascade and turbulent damping}
\label{subsec:FG}

The turbulent damping (FG) of self-generated Alfv\'en waves is due to
their interaction with a pre-existing background turbulence.  Such
turbulence may  be injected by astrophysical sources (see e.g.
\citealt{MacLow-2004}) with a turbulent velocity $v_{\rm turb}$ and
on scales, $L_{\rm inj}$, much larger than the CR Larmor radius.
For waves in resonance with particles with a given energy $E$, the
damping rate, that accounts for the anisotropy of the turbulent
cascade, has been derived by \cite{Farmer-2004,
Yan-Lazarian-2004-compressible} and reads
\begin{equation} \label{eq:Gfg}
    \Gamma_d^{\rm FG} = \left(\frac{v_{\rm turb}^3/L_{\rm inj}}{r_{\rm L}\, v_{\rm A}} \right)^{1/2},
\end{equation}
where $v_{\rm A}$ is the effective Alfv\'en speed felt by CRs, as defined
in Sec.~\ref{dampingaw}.
We take the turbulence as trans-Alfv\'enic at the injection scale,
namely $v_{\rm turb} = v_{{\rm A}, n}$ (at large scales waves are in the low wavenumber regime, where ions and neutrals are well coupled, as illustrated in Sec.~~\ref{dampingaw}).
This is the likely situation if
the turbulence is mainly injected by old SNRs, with a forward shock
becoming trans-sonic and trans-Alfv\'enic.
The FG damping rate is
shown in Fig.~\ref{fig:damping_WNM_WIM} for the WIM and WNM. 

In highly neutral media, such as the
the WNM, CNM and DiM, the background turbulence responsible for the FG damping  can be damped by ion-neutral friction  at a scale, $ l_{\rm min} = 1/k_{\rm min}$ \citep{Xu-2015-ApJ...810...44X, Xu-2016-ApJ...826..166X, Lazarian2016, Brahimi2020}. Correspondingly, there is a minimum particle energy, $E_{\rm min}$, such that $r_{\rm L}(E_{\rm min}) = l_{\rm min}$, below which the FG damping cannot affect the self-generated Alfv\'en waves \citep{Brahimi2020}:
\begin{equation}
 \frac{1}{l_{\rm min}} = L_{\rm inj}^{1/2}
\left(\frac{2\epsilon\nu_{in}}{v_{{\rm A}, n}}\right)^{3/2} \sqrt{1 + \frac{v_{{\rm A}, n}}{2\epsilon\nu_{in} L_{\rm inj}}}.
\end{equation}
In Fig.~\ref{fig:damping_WNM_WIM} the FG damping rate for the WNM is truncated at $E_{\rm min}$. 
In the WIM the cascade rate is found to be always larger than the ion-neutral damping rate and there is no $E_{\rm min}$.

\subsection{Non-linear Landau damping}

The non-linear Landau (NLL) damping is caused by  the interaction between the beat of
two Alfv\'{e}n waves and the thermal (at temperature $T$) ions in
the background medium. The damping rate for resonant waves is given by
(\citealt{Kulsrud-1978-nlld,  Wiener-2013-WIM-ion-neutral})
\begin{equation}
    \Gamma_d^{\rm NLL} = \frac{1}{2}\sqrt{\frac{\pi}{2}
    \left(\frac{k_{\rm B}\,T}{m_{\rm p}}\right)}
    \frac{I(k_{\rm res})}{r_{\rm L}},
\end{equation}
where 
$k_{\rm B}$ is the Boltzmann constant
and $I(k_{\rm res})$  is the wave energy density (see Sec.~\ref{sec:cr-prop} below for the definition) at the resonant wavenumber $k_{\rm res} = 1/r_{\rm L}$.

\section{Cosmic ray propagation in the proximity of SNRs}
\label{sec:cr-prop}

We consider the escape of CRs from a SNR and the subsequent propagation
in the source proximity.
The propagation region is assumed to be embedded in a turbulent
magnetic field, with a large scale ordered component of strength
$B_0$. CRs are scattered by Alfv\'en waves, which constitute a
turbulent magnetic field background of relative amplitude $\delta
B/B_0$, where $k$ is the wavenumber. We only consider waves that propagate along the uniform background field $B_0$. In the limit of $\delta
B/B_0 \ll 1$, which is the one relevant for the cases treated in this paper, the CR diffusion along field lines can be treated in the quasi-linear regime, with a
diffusion coefficient given by \cite{Berezinskii-1990} and \cite{Kulsrud-2005}
\begin{equation}\label{eq:D-QLT}
D(E)=\left.\frac{4\pi c\,r_{\rm L}(E)}{3 I(k_{\rm res})}\right|_{k_{\rm res}=1/r_{\rm L}} = \frac{D_{\rm B}(E)}{I(k_{\rm res})},
\end{equation}
where $c$ is the speed of light, $I(k_{\rm res}) = \delta B(k_{\rm res})^2/ B_0^2$ is the wave energy density  calculated
at the resonant wavenumber $k_{\rm res}=1/r_{\rm L}$,
and 
$D_{\rm B}(E) = (4\pi/3) c r_{\rm L}(E)$ is the Bohm diffusion coefficient. 
We also assume that the dominant source of Alfv\'enic turbulence
is produced by the CR resonant streaming instability. 

In our model we adopt the flux tube approximation for the CR transport  along $B_0$
\citep[see, e.g.,][]{Ptuskin-2008}, and we neglect the diffusion across field lines, which is suppressed in the $\delta
B/B_0 \ll 1$ regime (see e.g. \citealt{Drury-1983, Casse-2002-chaotoc-B}). Thus,  we do not address the perpendicular evolution of the flux tube \citep[see, e.g.,][]{Nava-2013-anisotropic} and any possible CR feedback on it and, in general, on the ISM dynamics \citep[see, e.g.,][]{Schroer-2020arXiv201102238S}.
Such one-dimensional model for the CR propagation is applicable for distances from the source below the coherence length, $L_{\rm c}$, of the  background 
magnetic turbulence, i.e. the scale below which the magnetic flux tube is roughly preserved \citep[see, e.g.,][]{Casse-2002-chaotoc-B}. 

When particles diffuse
away from the source at distances larger than 
$L_{\rm c}$, diffusion becomes 3-D and the CR density drops quickly. In the Galactic disk, $L_{\rm c}$ is estimated observationally and may range
from few pc to $\approx 100$ pc, depending on the ISM phase \citep[see, e.g.,][and references therein]{Nava-2013-anisotropic}.

We follow the approach proposed by \cite{Nava-2016-esc-I,
Nava-2019-esc-II}, and we determine: ({\it i}\/) the escape time of
CRs of  a given energy from the remnant; ({\it ii}\/) the time
dependent evolution of the CR cloud and of the self-generated
diffusion coefficient after escape; ({\it iii}\/) the time spent
by CRs in the source vicinity and the corresponding grammage.
We focus on the warm ionized and warm neutral phases of the ISM.
As shown in the following sections, the results depend significantly 
on the considered ISM phase, and
in particular on the amount and type of neutral and ionized atoms
in the background medium. 

\subsection{Transport equations and initial/boundary conditions}

The coupled CR and wave transport equations read 
(\citealt{Nava-2016-esc-I, Nava-2019-esc-II})
\begin{equation}\label{eq:CR-transp-time}
    \frac{\partial P_{\rm CR}}{\partial t} + v_{\rm A}\frac{\partial P_{\rm CR}}{\partial z}  = \frac{\partial}{\partial z}\left(\frac{D_{\rm B}}{I} \frac{\partial P_{\rm CR}}{\partial z}\right)
\end{equation}
and
\begin{equation}\label{eq:wave-transp-time}
    \frac{\partial I}{\partial t} + 
    v_{\rm A}\frac{\partial I}{\partial z} 
    = 2(\Gamma_{\rm CR} - \Gamma_d)I +Q.
\end{equation}
Here $v_{\rm A}$ is an effective Alfv\'en velocity that takes into
account the coupling between ions and neutrals (see
Sec.~\ref{dampingaw}), while $P_{\rm CR}$ is the partial pressure
of CRs of momentum $p$ normalized to the magnetic field pressure
\begin{equation}
    P_{\rm CR} = \frac{4\pi}{3}c\,p^4 \frac{f(p)}{B_0^2/8\pi}.
\end{equation}
The term $2\,\Gamma_{\rm CR}I$ gives the wave growth due to 
the CR streaming instability and can be expressed as
\begin{equation}
    2\Gamma_{\rm CR}I = - v_{\rm A}\frac{\partial P_{\rm CR}}{\partial z}.
\end{equation}
The term $\Gamma_d$ encompasses the relevant wave damping 
rates, which are described in detail in Sec.~\ref{sec:damp}.
The term $Q$ represents the possible injection of turbulence from
an external source, other than the CR streaming. Here it is taken
as $Q= 2\Gamma_{\rm d}I_0 $, where $I_0$ is 
a parameter such that typical values of the Galactic CR diffusion coefficient are
recovered at large distances from the source 
\citep[see, e.g.,][]{Strong-2007}. In this way, when the streaming instability
is not relevant, diffusion is regulated by the background turbulence.
Notice that in these equations we neglected adiabatic losses that arises when  Alfv\'en speed varies in space \citep[see, e.g.,][]{Brahimi2020}, since we are considering a homogeneous ISM around the SNR. 

The coordinate $z$ is taken along $B_0$ and $z
= 0$ refers to the centre of the CR source. The equations are solved
numerically with a finite difference explicit method, using the initial conditions
\be
    P_{\rm CR} = 
    \left\{
    \begin{array}{ll} 
     P^0_{\rm CR} & \mbox{if~}z< R_{\rm esc}(E), \\
     0            & \mbox{if~}z>R_{\rm esc}(E),  \nonumber
     \end{array} \\
     \right.
\ee
and
\be
    I = I_0\qquad\mbox{everywhere}.
\label{eq:Pcr-initial-cond}
\ee
Here $R_{\rm esc}(E)$ is the size of the region filled by CRs at
the time of escape, and $P^0_{\rm CR}$ is the initial CR pressure
inside this region.
The method used
to determine the escape radius and time for particles of energy $E$
is described in detail in \cite{Nava-2016-esc-I, Nava-2019-esc-II}.
As for the initial condition for the waves, it is also possible to
choose $I \gg I_0$ for $z< R_{\rm esc}(E)$ in order to mimic Bohm
diffusion inside the source, as proposed by \cite{Malkov-2013}.
However, as discussed by \cite{Nava-2019-esc-II}, different choices
of $I$ inside the source have little impact on the final solution.
As boundary conditions we impose a symmetric CR distribution at $z=0$, and 
\be
P_{\rm CR} =0, \qquad I=I_0 \qquad \mbox{~at~} %z=\infty.
z \gtrsim L_{\rm c}.
\label{eq:Pcr-boundary-cond}
\ee
The 1-D model used in this paper is valid only up to a distance from
the source given by the coherence length $L_{\rm c}$ of the magnetic
field. At larger distances the diffusion becomes 3-D and the CR
density quickly drops, a behavior that can be described in terms
of a free escape boundary at $L_{\rm c}$. The value of $L_{\rm c}$ is constrained
from observations to be $\lesssim 100$~pc but its value is matter
of debate and is likely phase- and position-dependent in our galaxy.
We take the
free escape boundary at $z \gtrsim L_{\rm c}$, and we check that
this assumption is not significantly affecting the results, for instance changing $L_{\rm c}$ to $10\,L_{\rm c}$.

\section{Results}
\label{sec:res}

In what follows we discuss the release of CRs of energy $E$ from a
SNR, giving an estimate of the age and radius of the source at the
moment of escape.  Moreover, we investigate the propagation of
runaway CRs in the source proximity, with particular focus on the
CR spectrum and self-generated diffusion coefficient.  Finally, we
estimate the residence time of CRs  in the source proximity and we
infer general implications for the grammage accumulated in that
region.

We treat these aspects in the cases of a WIM or a WNM surrounding
the remnant, with special emphasis on the role played by the
ion-neutral damping of Alfv\'en waves.  
  
In our calculations we assume that: ({\it i}\/) runaway CRs have a total
energy $E_{\rm CR} = 10^{50}$~erg, a power-law spectrum in energy
between 1~GeV--5~PeV with spectral index $\gamma = -2.0$; and ({\it
ii}\/) the typical ISM diffusion coefficient is $D(E) = D_0(E/10\,{\rm
GeV})^{0.5}$ with $D_0 = 10^{28}$~cm$^2$~s$^{-1}$ \citep[see, e.g.,][]{Strong-2007}.

\subsection{Escape of cosmic rays: source age and radius}

The age and radius of escape of CRs of energy $E$ is estimated as
follows \citep{Nava-2016-esc-I, Nava-2019-esc-II}: we define the
half-time, $t_{1/2}(E, R)$ of a CR cloud as the time after which
half of the CRs of a given energy, initially confined within a
region of size $R$, have escaped that region. The evolution of such
CR cloud is studied by solving Eq.~(\ref{eq:CR-transp-time}) and
Eq.~(\ref{eq:wave-transp-time})  with initial conditions
given by Eq.~(\ref{eq:Pcr-initial-cond}) and boundary conditions given by
Eq.~(\ref{eq:Pcr-boundary-cond}).

The radius of a SNR expanding in a homogeneous medium of density
$n$ can be estimated as \citep{TM-1999}
\begin{equation}
    R_{\rm SNR}(t) = 0.5 \left(\frac{E_{51}}{n}\right)^{1/5}\left[1-\frac{0.09 M_{{\rm ej}\odot}^{5/6}}{E_{51}^{1/2}n^{1/3}t_{\rm  kyr}}\right]^{2/5}t_{\rm  kyr}^{2/5}~\mbox{pc},
\end{equation}
where $E_{51}$ is the supernova explosion energy in units of
$10^{51}$~erg, $n$ is the total density of the ambient ISM in $\rm
cm^{-3}$, $M_{{\rm ej}\odot}$ is the mass of the supernova ejecta
in solar masses, and $t_{\rm kyr}$ is the SNR age in kyr. This
equation is valid in the adiabatic phase of the expansion, which
starts after $\approx 86\,M_{{\rm ej}\odot}^{5/6}E_{51}^{-1/2} n^{-1/3}$~yr. Here we take $E_{51} = 1$ 
and $M_{{\rm ej}\odot} = 1.4$, while
the gas density depends on the medium and is reported in
Table~\ref{table:ISM}. At earlier times an approximate expression
for the free expansion phase has to be used \citep{Chevalier-1982}.
The adiabatic phase stops at roughly $\approx 1.4\times10^4
E_{51}^{3/14}n^{-4/7}$~yr, when the radiative phase starts
\citep{Cioffi-1988}. At this epoch we also assume that the acceleration
of CRs becomes ineffective and that all CRs are instantly released \citep[the validity  of the assumption of instant release may depend on the conditions in the shock precursor, as discussed by][]{Brahimi2020}.

The escape time, $t_{\rm esc}(E)$, of particles with energy $E$ is
estimated as the time such that $t_{1/2}(E, R_{\rm SNR})$ equals
the age of a SNR of radius $R_{\rm SNR}$. This is the timescale
over which waves can grow. The dependence of $t_{1/2}(E, R)$ on
the initial radius $R$ at fixed energy $E$, and varying the background
diffusion coefficient and the CR spectral index, has been extensively
explored by \cite{Nava-2016-esc-I, Nava-2019-esc-II}, and we refer
the reader to these works for a detailed discussion. Here we briefly
summarize the most relevant points. At small $R$ the CR gradient
is large and the amplification of waves is very effective. In this
regime, the NLL damping, that scales with the wave energy density $I(k)$,
can play an important role and may dominate over ion-neutral and FG damping
at small enough radii. At intermediate $R$, ion-neutral damping dominates
in most cases. In fact, as shown in Fig.~\ref{fig:damping_WNM_WIM},
the ion-neutral damping rate is larger than the FG rate at least up to
particle energies of $\sim 10$~TeV, with the only exception
of a WIM with fully ionized hydrogen and $\sim 10\%$ of neutral
helium. At large $R$, the CR gradient is reduced and the streaming
instability is less effective. In this case the expansion of the
CR bubble is determined by the background turbulence and the test
particle limit is recovered, with $t_{1/2} \propto R^2/D_0$.

\begin{figure*}
%\centering {
\includegraphics[width=\columnwidth]{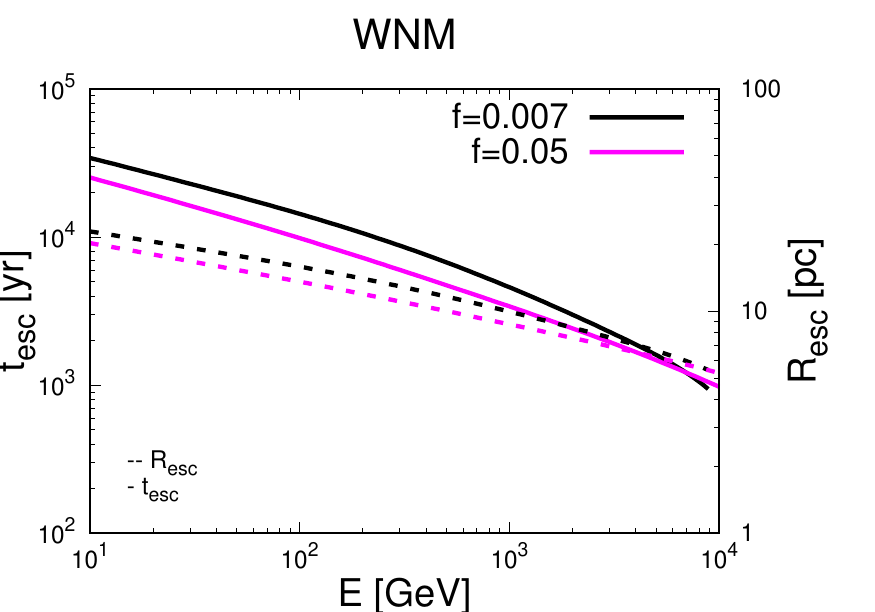}
\includegraphics[width=\columnwidth]{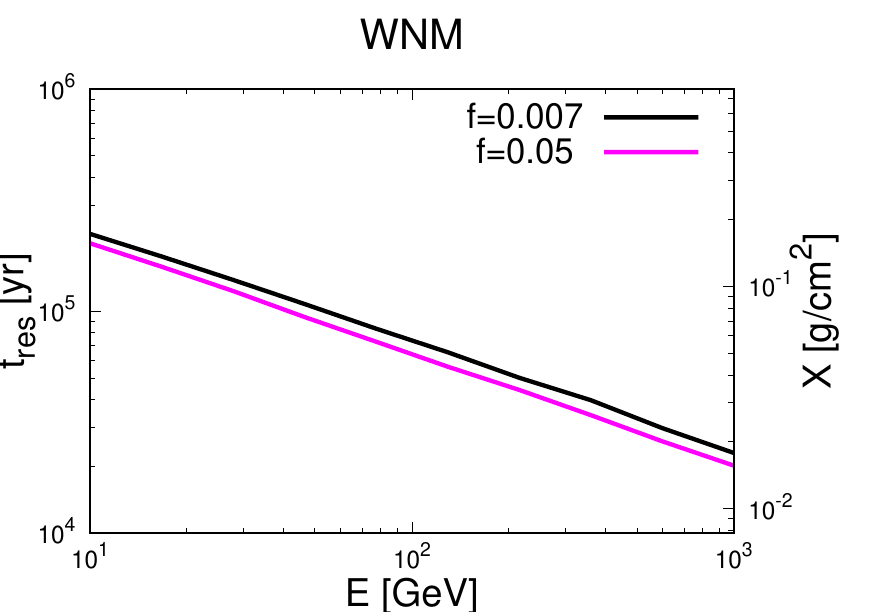}
%}
\caption{Case of WNM. \textit{Left panel:} SNR age and radius at the time
of escape of CRs, as a function of the CR energy. \textit{Right
panel:} Residence time of CRs in a region of $\sim 100$~pc around
the source, as a function of the CR energy, and the corresponding
grammage. Different colors are used for different values of the
hydrogen ionization fraction $f$, while the helium-to-hydrogen ratio
$\chi = 0.1$. The parameters of the WNM are listed in
Table~\protect\ref{table:ISM}.}
\label{fig:ESC-RES-WNM}
\end{figure*}

\begin{figure*}
%\centering {
\includegraphics[width=\columnwidth]{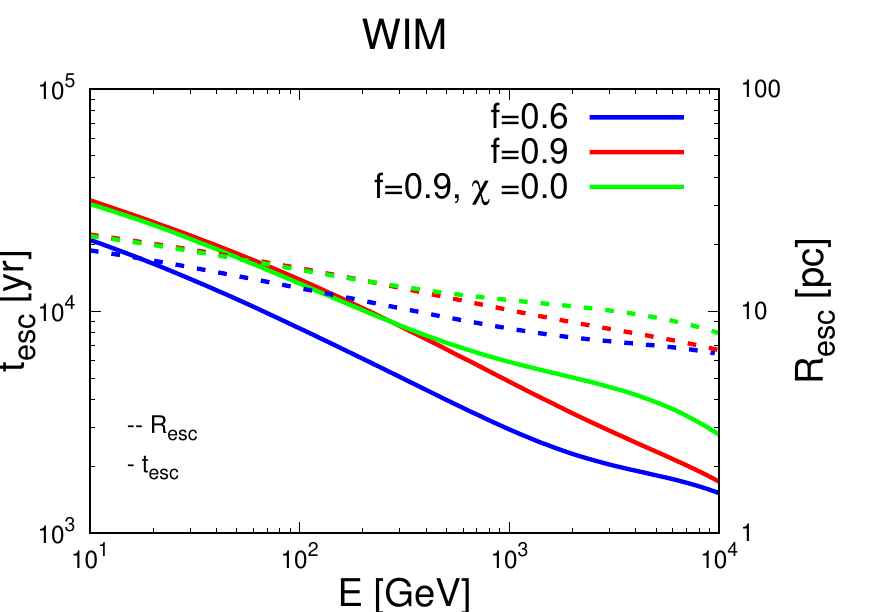}
\includegraphics[width=\columnwidth]{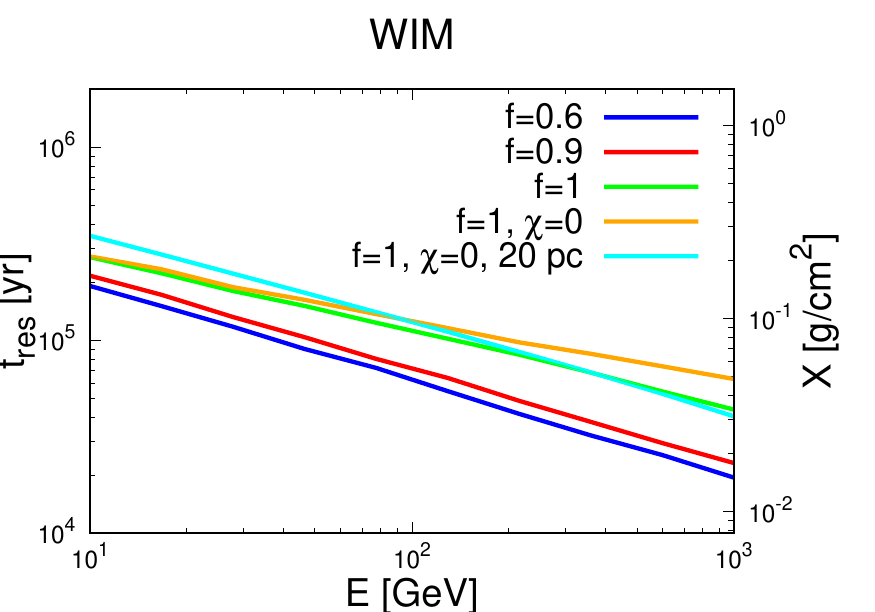}
%}
\caption{Case of WIM. \textit{Left panel:} SNR age and radius at the time
of escape of CRs, as a function of the CR energy. \textit{Right
panel:} Residence time of CRs in a region of $\sim 100$~pc around
the source, as a function of the CR energy, and the corresponding
grammage. The cyan curve is computed by assuming that particles
of all energies escape at SNR radius of 20~pc, as assumed by
\protect\cite{Dangelo-2016-grammage}. Different colors are used
for different values of the hydrogen ionization fraction $f$ and
helium-to-hydrogen ratio $\chi$. The  parameters of the WIM are
listed in Table~\protect\ref{table:ISM}.}
\label{fig:ESC-RES-WIM}
\end{figure*}

Fig.~\ref{fig:ESC-RES-WNM} and Fig.~\ref{fig:ESC-RES-WIM} show the SNR
age and radius at the time of escape of CRs, as a function of the
CR energy,in the case of a WNM and WIM respectively.  We confirm
the findings by \cite{Nava-2016-esc-I, Nava-2019-esc-II} and the
qualitative results that particles with higher energy escape earlier
than the low energy particles (see, e.g., \citealt{Gabici-2011-escape}).

In the case of the WIM, $R_{\rm esc}$ and $t_{\rm esc}$ tend to
decrease with a larger neutral density.  The effect of ion-neutral damping
is especially visible in the energy range $\rm\sim 1$--10~TeV,
where the ion-neutral damping rate starts to decrease from a roughly constant
value at different particle energies, depending on the value of the
ionized hydrogen fraction $f$ and on the helium-to-hydrogen ratio
$\chi$, as shown in Fig.~\ref{fig:damping_WNM_WIM}. The drop of
$\Gamma_d^{in}$ corresponds to an increase of $t_{\rm esc}$,
and therefore to a better CR confinement.

The situation is more subtle in the case of the WNM. In fact, here
the confinement appears to be slightly more effective for a smaller
value of $f$, a result opposite to what happens in the WIM.  This
can be explained taking into account that, while the $\Gamma_d^{\rm
in}$ is practically the same at energies below $\rm\sim 10$~TeV
for $f\sim 7\times10^{-3}-5\times10^{-2}$, the effective Alfv\'en
speed felt by CRs, which at these energies is roughly that of ions,
as illustrated in Sec.~\ref{sec:damp}, is a factor of $\sim 2$--3
larger for $f=7\times10^{-3}$. This reflects on an increase by the
same amount of the FG damping rate (above the cut-off energy, see Sec.~\ref{subsec:FG}), as shown in
Fig.~\ref{fig:damping_WNM_WIM}, but also of the growth rate term,
which is proportional to $v_{\rm A}$. On the other hand, at energies
in the range $\rm\sim 10$~GeV--1~TeV, the FG damping is always
subdominant compared to the ion-neutral damping, and the net effect of this
change of $v_{\rm A}$ is a slightly better confinement at smaller
$f$ as a result of an enhanced wave growth rate.

Also in the case of a WIM the effective $v_{\rm A}$ is a bit larger
at smaller $f$ below $\sim 1$~TeV, but here the increase is
only by a factor of $\sim 1.2$ and cannot overcome the effect of
ion-neutral damping.

\subsection{Cosmic ray spectra and diffusion coefficient}

\begin{figure*}
\includegraphics[width=\textwidth]{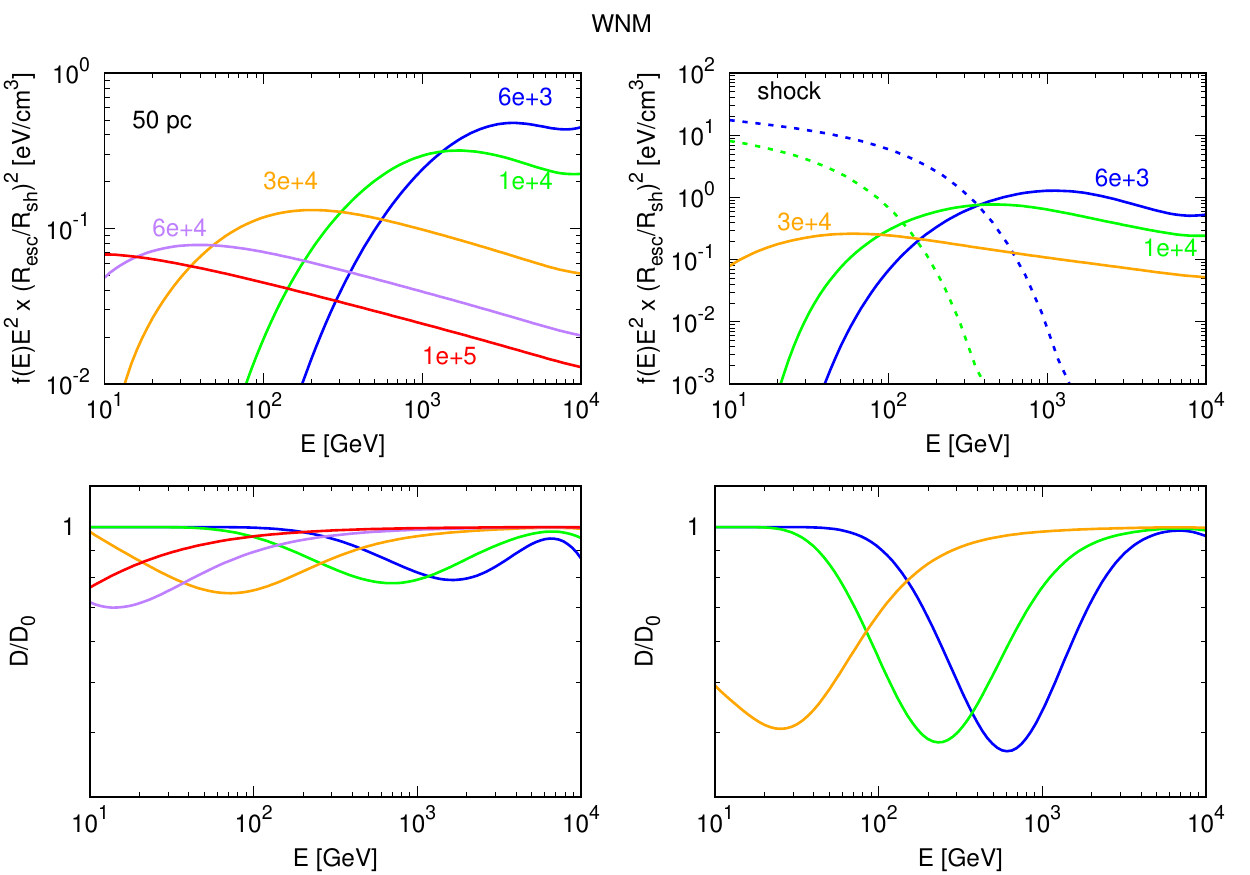}
\caption{Case of WNM ($f=0.05$, $\chi=0.1$): CR spectrum and ratio 
$D/D_0$ as a function of energy, where $D_0(E) = 10^{28} (E/{\rm 10\,GeV})^{0.5}$~cm$^2$~s$^{-1}$. 
The different colors refer to 
different ages in yr (as marked). The left panels refer to a 
location at 50~pc from the center of the SNR, while the right panels 
refer to the shock position. In this case also the spectrum of CRs
still confined in the accelerator is shown (dashed lines).}
\label{fig:WNM-spectra}
\end{figure*}

\begin{figure*}
 \centering
\includegraphics[width=\textwidth]{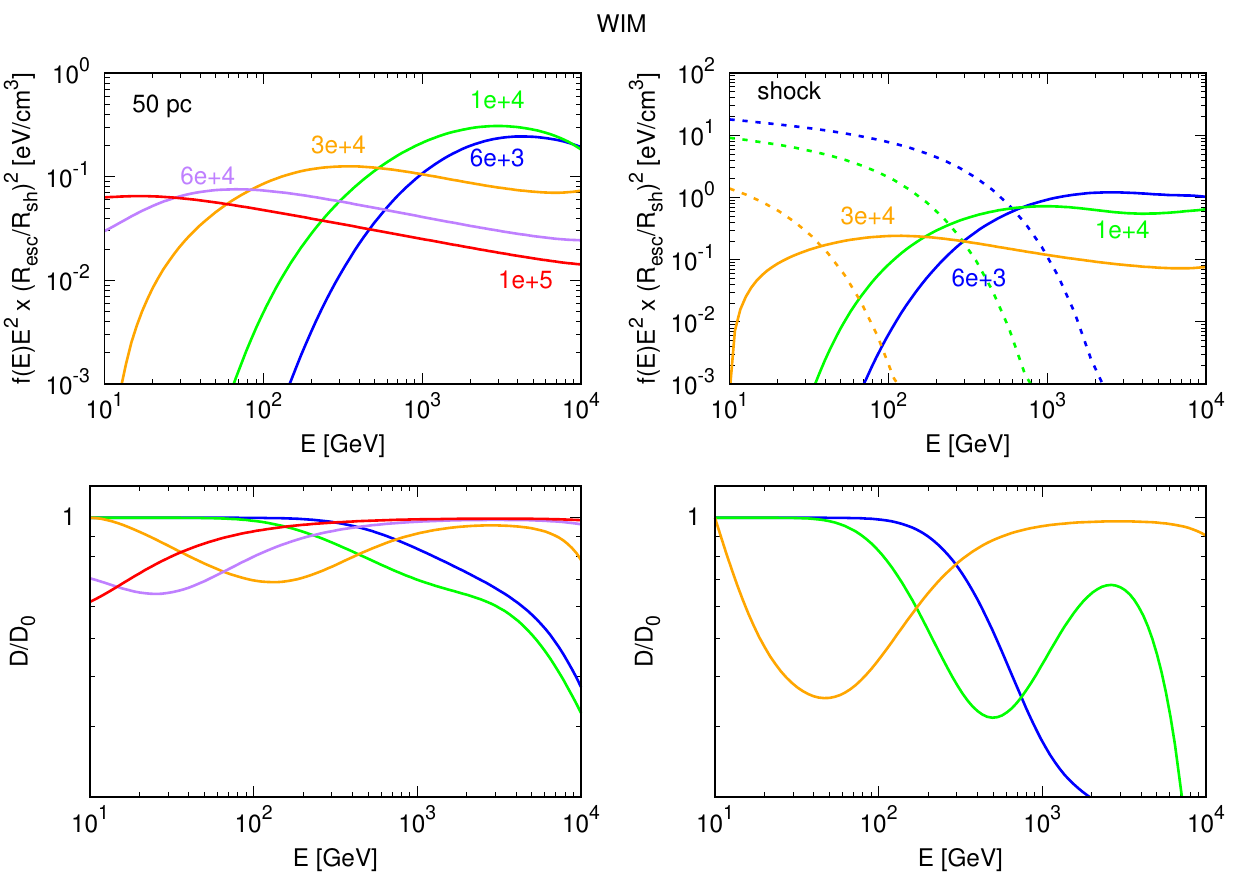}
\caption{Case of WIM ($f=0.9$, $\chi=0.1$): CR spectrum and ratio $D/D_0$
as a function of energy, where $D_0(E) = 10^{28} (E/{\rm 10\,GeV})^{0.5}$~cm$^2$~s$^{-1}$. The different colors refer to different
ages in yr (as marked). The left panels refer to a location at 50~pc
from the center of the SNR, while the right panels refer
to the shock position. In this case also the spectrum of CRs still
confined in the accelerator is shown (dashed lines).}
\label{fig:WIM-spectra}
\end{figure*}

The spectrum of runaway CRs and the corresponding self-generated
turbulence depend significantly on the distance from the source and
on the time. In Fig.~\ref{fig:WNM-spectra} and
Fig.~\ref{fig:WIM-spectra} we show the spectrum of CRs that
have already escaped to a distance of 50 pc from the centre of the
remnant at different times, from $t=6\times 10^3$~yr to $t=
10^5$~yr, and for the WNM ($f=0.05$, $\chi=0.1$)  and the WIM ($f=0.9$, $\chi=0.1$), respectively. To remove the effect of the
simple 1-D geometry that we are adopting we multiply $P_{\rm CR}$
by $R_{\rm esc}^2(E)/R^2_{\rm SNR}(t)$. For each case we also show
the ratio between the self-generated and background diffusion
coefficients.

In the same figures we also show the spectrum and diffusion coefficient
at the position of the shock of CRs that have escaped at earlier
times (and are considered as decoupled from the accelerator) as
well as the spectrum of particle still confined in the accelerator,
at different ages of the remnant from $t=6\times 10^3$~yr to $t=
3\times 10^4$~yr. The spectrum of the confined particles is estimated
by assuming that the remnant provides $\approx 10^{50}$ erg in CRs
with a $\propto E^{-2}$ spectrum that extends from $\sim 1$~GeV 
to $\sim 5$~PeV, with an exponential cutoff at the energy
of the particles that escape at the considered age. The SNR is
assumed to stop accelerating particles at the onset of the radiative
phase, which for the typical density of the WIM and WNM takes
place at $t_{\rm rad}\approx 3\times 10^4$~yr. The corresponding
shock radius is $R_{\rm rad}\approx 22$~pc. In all cases, at each
time and location, the spectrum exhibit a sharp rise at low energies
and a peak followed by a nearly power-law like behaviour at higher
energies.

Since high energy particles escape earlier and diffuse faster, they
reach a given distance at earlier times, as shown for the spectra
at 50~pc, where the peak moves at lower energies with time.  At
energies lower than the peak, particles have not yet reached that
position, giving the sharp rise. At larger energies, particles have
diffused over a bigger distance, giving a spectrum steeper than the
spectrum released from the SNR, here assumed to be $\propto E^{-2}$.

A similar behaviour is observed with the spectra at the shock. Here the
position at which the spectra are shown varies with time, since it
is given by the shock radius at a given age. This radius is always
$< 50$~pc for the cases illustrated here. Comparing the spectra
at the shock and at 50~pc for a given age, it is evident that the
peak is at lower energy in the former case. This again is the result
of an earlier escape and faster diffusion of high energy particles.

As for the $D/D_0$ ratio as a function of energy, it is possible
to infer from the plots that, at a given time, the diffusion
coefficient is equal to its background value at low energies, where
particles have not yet escaped or not yet reached the position where
$D$ is calculated, and at high energies, where the turbulence
produced by such particle is being damped. At intermediate energies,
and roughly corresponding to the peak in the spectrum, the diffusion
coefficient is suppressed compared to $D_0$ due to CR streaming instability.
The energies at which the suppression is evident move to lower
values with time.

A more complex situation can be observed in some cases, for instance
at $t=10^4\, \rm yr$ at the shock and in the TeV range, for the
WIM. Here $D$ starts to deviate from $D_0$ at $\rm \sim 50$~GeV,
then, above $\rm \sim 500$~GeV, $D$ begins to approach again $D_0$,
but above $\rm \sim 3$~TeV the level of turbulence increases again,
with the result of a suppression of $D$. This is due to the energy
dependence of the escape radius, which becomes steeper above few
TeV. This results in the fact that, above that energy, the escape
radius becomes small enough as to effectively excite the streaming
instability. In fact the density of runaway CRs is $\propto 1/R_{\rm
esc}^2$. The more efficient self-confinement at this high energies
also reflects in a hardening of the spectrum (see also
\citealt{Nava-2019-esc-II} for a discussion).

\subsection{Residence time and grammage}

The grammage 
%possibly
accumulated by CRs close to their sources,
and its relevance compared to the grammage accumulated while diffusing
in the whole Galaxy, is related to the residence time in the proximity of the source.  A formal determination  of the grammage should be done
by solving the CR transport equation for nuclei with the inclusion
of spallation contributions \citep[see, e.g.,][]{Berezinskii-1990,
Ptuskin-Soutoul-1990}. 

Here we adopt the following approach: we assume that $N_0$ particles of energy $E$ are instantly injected by a source located in a region of constant gas density $\rho=\mu m_p n$ and are subject to free escape at a boundary located at a distance $L_{\rm c}$ from the source.
At a given time $t$, the total number of particles that are still in the region is $N_{\rm in}(t)$ while the number of  escaped  particles is $N_{\rm esc} = N_0 - N_{\rm in}(t)$. Particles  that cross the boundary at time $t$ have accumulated the grammage $X(t)=\rho c t$ and the average grammage gained  by all escaped particles from $t=0$ to $t=\infty$ is given by
\begin{align}
    \langle X \rangle & = \frac{1}{N_0} \int_0^{\infty} \rho c t \frac{dN_{\rm esc}}{dt}\, dt \\ \nonumber
    & =\frac{\rho c}{N_0}\left[ \int_0^{\infty} N_{\rm in}(t)\, dt - \lim_{t \longrightarrow \infty} (t\, N_{\rm in}(t))   \right]\\ \nonumber
    & = \rho\, c\, \tau_{\rm res}.
\end{align}
%Here $\rho = 1.4\, m_p n $, where $n$ is the total gas density, and  the 
The residence time $\tau_{\rm res}$ is given by
\begin{equation}
    \tau_{\rm res} = \frac{1}{N_0}\left[ \int_0^{\infty} N_{\rm in}(t)\, dt - \lim_{t \longrightarrow \infty} (t\, N_{\rm in}(t))   \right].
\end{equation}
The number of particles with a given energy contained in the region at a given time can be calculated from the CR pressure (see Sec.~\ref{sec:cr-prop}):
\begin{equation}
    N_{\rm in}(t)= \frac{N_0}{R_{\rm esc}P_{\rm CR}^0}\int_0^{L_{\rm c}} P_{\rm CR}(z, t, E)\, dz,  
\end{equation}
where we took into account that CRs on energy $E$ are initially released from a region of size $R_{\rm esc}(E)$.

Fig.~\ref{fig:ESC-RES-WNM} and \ref{fig:ESC-RES-WIM} show the
residence time of CRs in a region of $\sim 100$~pc around the
remnant, for the WNM and the WIM, respectively, and for different
values of the neutral fraction, and the corresponding grammage.
Even in the (not very plausible) case of a fully ionized WIM ($f=1.0$,
$\chi=0.0$), the grammage is nearly two orders of magnitudes smaller
than that accumulated in the disk (see e.g. \citealt{Jones-2001-grammage,
Gabici-2019-review-con-altri}). This is due to the action of the
ion-neutral and FG damping.  The presence of a $\sim 10$\% of neutral
helium (which is less effective at wave damping compared to hydrogen)
noticeably reduces the CR residence time compared to the case of a
fully ionized background medium, above few tens GeV.

These results are at odds with what previously suggested by
\cite{Dangelo-2016-grammage}, namely that ion-neutral damping is totally
negligible in the case of only neutral helium, since the c.e. cross
section is very small. However, as shown in Sec.~\ref{sec:damp},
not only c.e. but also m.t. has to be taken into account in the ion-neutral
damping.  In fact, \cite{Dangelo-2016-grammage} find a grammage
nearly a factor of ten larger than what we find. This is in part
due to their assumption of ion-neutral damping of waves with neutral helium.
In addition, they assume a 20\% acceleration efficiency, while we
assume 10\%, which enhances the effect of streaming instability,
and they do not include the FG damping. Finally,
\cite{Dangelo-2016-grammage} assume that CRs at all energies are
released when the SNR radius is 20~pc, a scenario implying a smaller
release radius at low energies compared to our calculation, which
translates in an enhancement of the streaming instability (and the
CR confinement) at energies below $\approx 1$~TeV, as shown in
Fig.~\ref{fig:ESC-RES-WIM}.

Any addition of neutral hydrogen compared to the case of only neutral
helium further reduces the confinement time.  Correspondingly, the
CR grammage accumulated in the source proximity results to be well
below that inferred from observations. A similar result was found
by \cite{Nava-2019-esc-II} for the HIM, where the residence time
tends to be larger than in a partially ionized medium, but the ISM
density is lower ($\sim 0.01$~cm$^{-3}$).

\section{Conclusions}
\label{sec:conc}

We followed the method and the setup proposed by \cite{Nava-2016-esc-I,
Nava-2019-esc-II} to investigated the escape of CRs from SNRs
embedded in a WNM and WIM, and the CR self confinement in the source
proximity. Our main objective was to determine 
whether, in realistic situations, the grammage accumulated by CRs in the
source region could become comparable to that inferred from observations. \sout{If this was the case,} In this case,
the standard picture in which CR secondaries are produced during
the whole time spent by cosmic rays throughout the disk of the
Galaxy, should be profoundly revisited.

We confirm the results found by \cite{Nava-2016-esc-I,
Nava-2019-esc-II} that CRs escaping from SNRs drive the excitation of Alfv\'en waves through the resonant CR streaming instability,  which results in a suppression of the diffusion coefficient and in the CR self-confinement in the source region. The SNR radius at which CRs of a given energy leave the source, $R_{\rm esc}(E)$, results to be a decreasing function of the energy, and regulates the CR streaming instability through a dilution factor $\propto
1/R_{\rm esc}^2$. 

We found that the growth of self-generated Alfv\'en waves, and consequently the residence time of CRs in the source region, is significantly limited by several damping processes,  especially by the  FG and ion-neutral damping. In particular, for the ion-neutral damping of Alfv\'en waves we
have used up-to-date damping coefficients, based on accurate experimental or theoretical determinations of the momentum
transfer and charge exchange cross sections between ions and neutrals
of different species.

In the  1-D geometry adopted in our calculation a suppression of the diffusion coefficient is  found within a distance form the source of the order of the magnetic field coherence length $L_{\rm c} \approx 50-100$ pc,  for a timescale that can be as large as $\sim 10^5\/ \rm yr$ at $\sim 10$~GeV. A smaller value for $L_{\rm c}$ would make the CR propagation to become 3-D closer to the remnant, thus reducing the residence time compared to our results.

We conclude that ion-neutral damping strongly limits the CR grammage that can be accumulated in the source region. Under the conditions typically met in the WIM and WNM of the Galactic disk, the CR source grammage is found to be negligible compared to that 
inferred form observation. A similar result was found for the HIM by \cite{Nava-2019-esc-II}, where the CR residence time is typically  larger than in a partially ionized medium but the ISM density is more than a factor ten  smaller.
This  makes alternative scenarios for the interpretation of quantities such as the B/C ratio, in which an important contribution to the production of secondaries comes from  the source region, less attractive.

Our results on  the residence time (and grammage) may  change with the inclusion of  the contribution of other sources of turbulence to the CR confinement. In particular, it has recently been suggested that 
\sout{an other} another  CR-induced instability, the so called \textit{non-resonant streaming instability}, which is mediated by the CR current and plays a crucial role in the acceleration of CRs \citep{Bell-2004}, may significantly enhance the CR self-confinement in the source region above $\sim $ TeV energies \citep{Schroer-2020arXiv201102238S}. Further investigations are thus needed in order to firmly establish the \sout{importance} amount of CR grammage accumulated in the vicinity of sources by very-high energy particles.

\bibliographystyle{mnras.bst} 
\bibliography{biblio}

\appendix
\section{Collisional coefficients}

\subsection{Collisions of H$^+$ with H and He atoms}
\label{subsec:coll-HH-HHe}

In the WNM and the WIM the relevant damping processes for Alfv\'en waves are 
collisions of H$^+$ with either H or He atoms. 
The case of H$^+$-- H collisions is special because the proton
elastically scattered by the H atom is indistinguishable from the
recoiling proton produced by charge exchange (\citealt{ks99, gks05, s08, s16}). Therefore, the elastic scattering and the charge
exchange channels for collisions of H$^+$ with H, as for any collision
between ions and their parent gas, cannot in general be separated.
Only at collision energies $E_{\rm cm}\gtrsim 1$~eV
can the forward elastic and the backward charge
exchange 
be approximately separated. In this
limit, $\sigma_{\rm mt}\approx 2\sigma_{\rm ce}$ \citep[][]{d58},
a frequently used approximation (see e.g. \citealt{Kulsrud-1971}).

\begin{figure}
\includegraphics[width=\columnwidth]{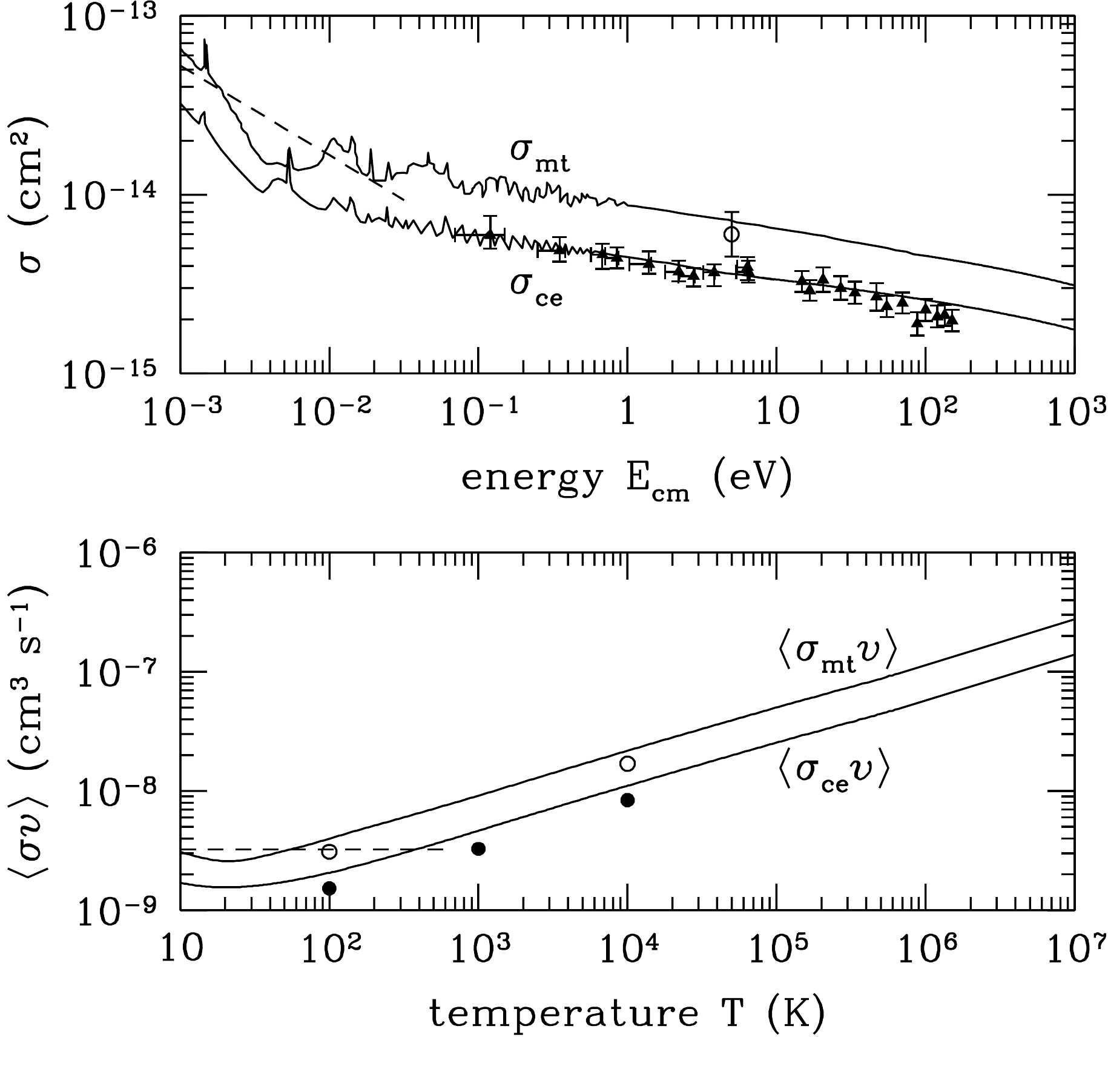}
\vspace*{-5mm}
\caption{
{\em Top panel:} cross sections for collisions of H$^+$
with H vs. collision energy in the center-of-mass
frame $E_{\rm cm}$. Momentum transfer cross section, from
\protect\cite{ks99}, \protect\cite{gks05}, and \protect\cite{s08}; charge exchange cross
section, from \protect\cite{hb91}, and \protect\cite{s08}; {\em dashed line:}
Langevin m.t. cross section. Experimental data for the c.e. cross section:
\protect\cite{n82} ({\em filled triangles}). The {\em empty circle}
shows the value of the m.t. cross section measured by
\protect\cite{b71}
{\em Bottom panel:} Collisional rate coefficient
for momentum transfer and charge exchange. The {\em filled} and {\em empty circles} show the rate coefficient adopted by \protect\cite{Kulsrud-1971} and \protect\cite{Zweibel-1982-ion-neutral}, respectively.
}
\label{ph}
\end{figure}

Fig.~\ref{ph} shows the m.t. cross section for H$^+$-- H collisions
(\citealt{ks99, gks05}), and the corresponding rate coefficient.
The figure also shows the m.t. cross sections and rate coefficients
derived from either charge exchange and elastic scattering in the
distinguishable particle approach (\citealt{s08, s16}), and the
experimental determination of the m. t. cross section
obtained by \cite{b71} by measuring the attenuation of Alfv\'en
waves propagating in a partially-ionized hydrogen plasma at $E_{\rm
cm} \approx 5$~eV. Also shown in Fig.~\ref{ph} are the rate
coefficients used by \cite{Kulsrud-1971} 
and \cite{Zweibel-1982-ion-neutral}, and frequently adopted
also for other species 
by scaling the collision
rate with the ratio $m_n/m_i$ of the neutral and ion masses (see e.g. \citep{Zweibel-1982-ion-neutral}).
As shown in the following, this
approximation is not generally correct, since the rate coefficients in the case of other species exhibit a different dependence on the temperature compared to the  H$^+$-- H collisions case.

\begin{figure}
%\begin{center}
\includegraphics[width=\columnwidth]{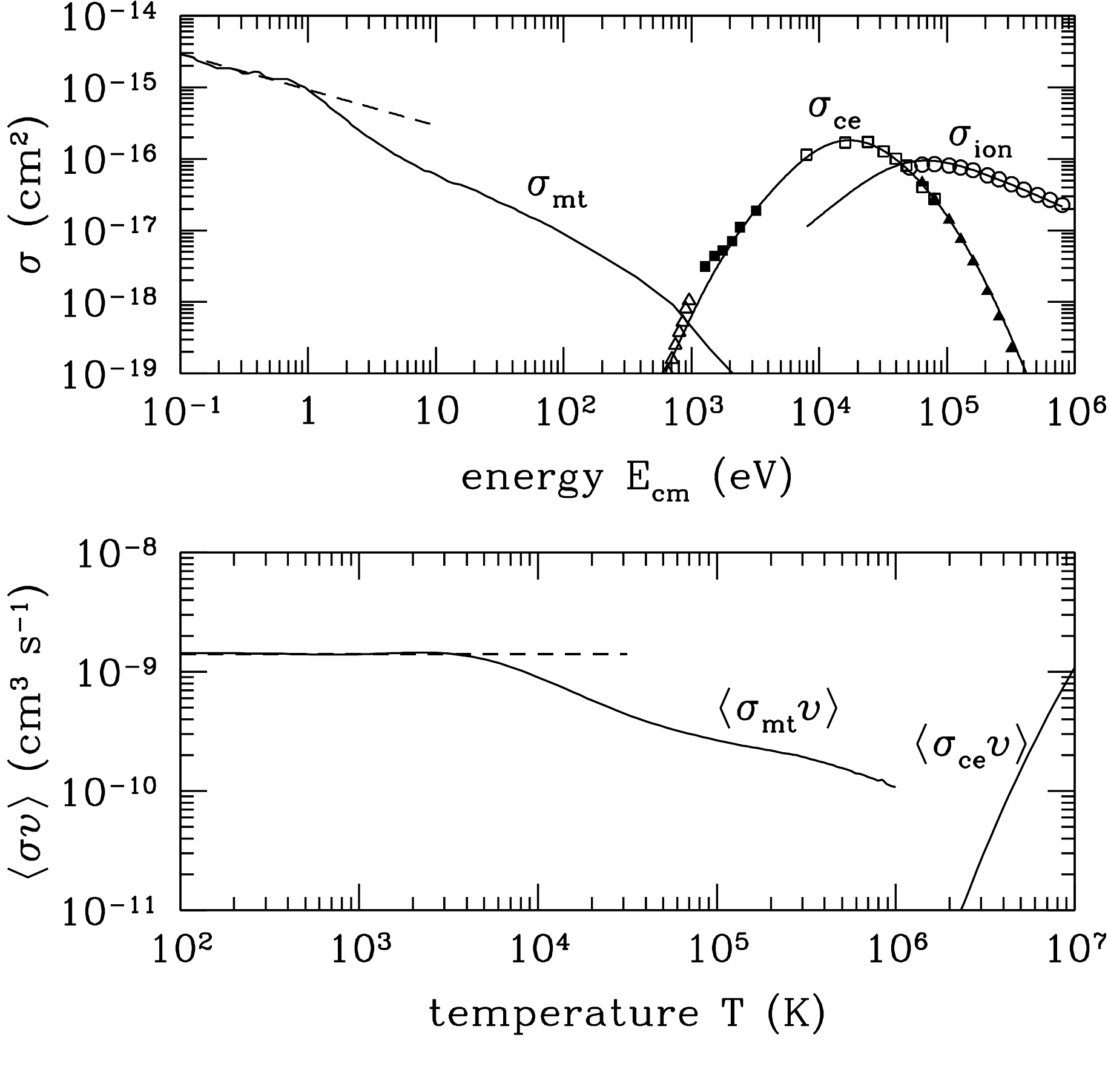}
\vspace*{-5mm}
\caption{
{\it Top panel:} cross sections for collisions of H$^+$
with He vs. collision energy in the center-of-mass
frame $E_{\rm cm}$. Momentum transfer cross section, from
\protect\citet{ks99, ks06}; Langevin m.t. cross
section ({\em dashed line}); charge exchange cross section,
from \protect\cite{l18}; and ionization cross section, from \protect\cite{sg85}. Experimental data for the c.e. 
cross section: \protect\cite{m81}
({\em empty triangles}), \protect\cite{sg85} ({\em filled
triangles}), \protect\cite{s89} ({\em empty squares}), and
\protect\cite{k11} ({\em filled squares}). Experimental data for the 
ionization cross section: \protect\cite{sg85} ({\em empty circles}). 
{\it Bottom panel:}
Collisional rate coefficient for Langevin m.t. ({\it dashed line}), m.t. and c.e. ({\it solid lines}).
}
\label{phe}
\end{figure}

Fig.~\ref{phe} shows the m.t. (\citealt{ks99, ks06}) and c.e.
\citep{l18} cross sections for H$^+$-- He collisions, and the
corresponding rate coefficients.
Charge exchange
contributes to the transfer of momentum above collision energies
$\sim 10^3$~eV, corresponding to relative velocities of $\sim
500$~km~s$^{-1}$, much larger than typical thermal or Alfv\'en
speeds in the WNM and WIM.

\subsection{Collisions of C$^+$ with H and He atoms}

\begin{figure}
\includegraphics[width=\columnwidth]{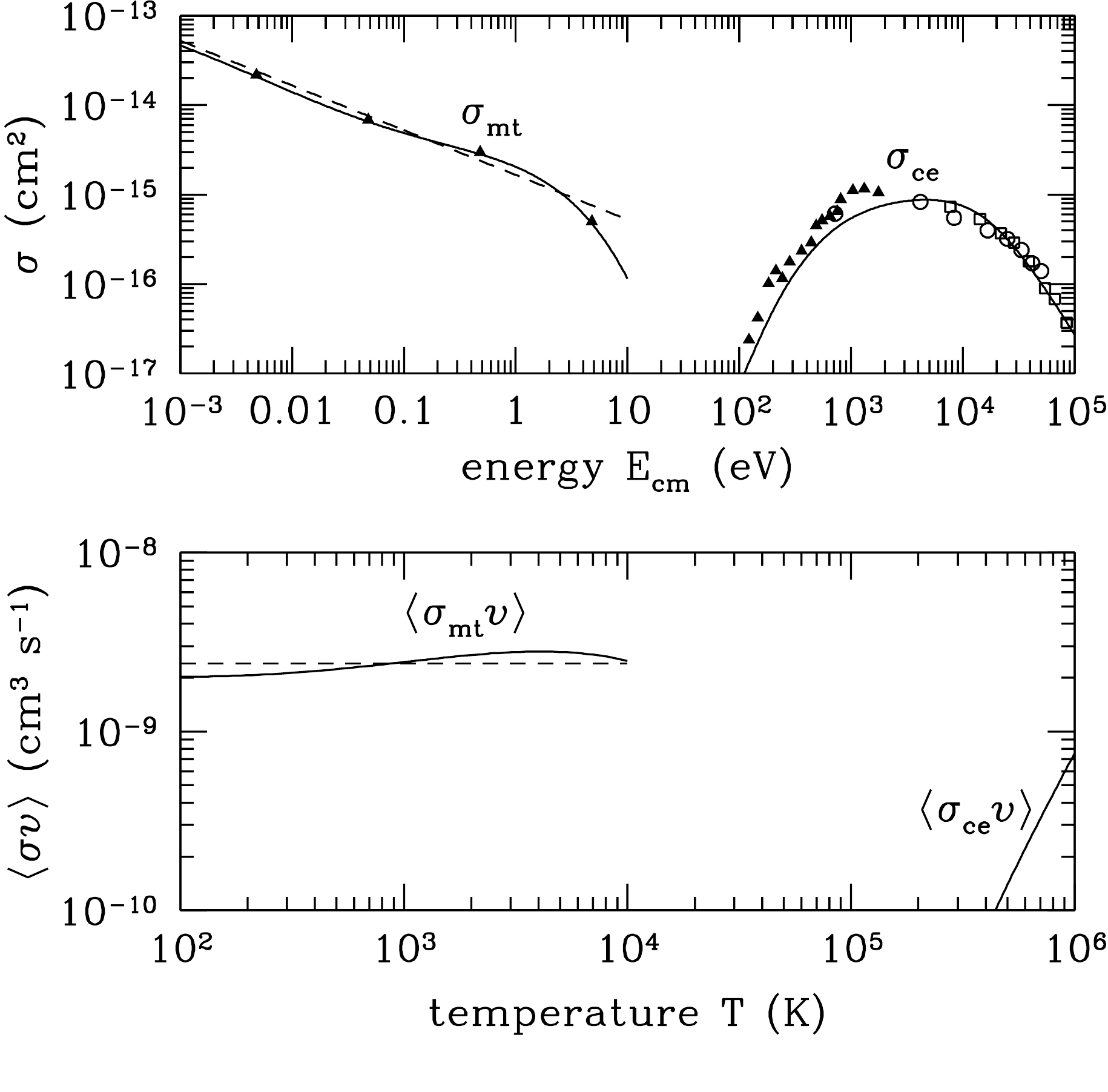}
\vspace*{-5mm}
\caption{{\it Top panel:} cross sections for collisions of C$^+$
with H vs. collision energy in the center-of-mass
frame $E_{\rm cm}$: m.t. cross
section computed by \protect\cite{fpdf95} ({\em filled triangles}); 
fit, from \protect\cite{pg08} ({\it solid line});  Langevin m.t.
cross section ({\it dashed line}).  Experimental data for the c.e.
cross section: \protect\cite{Phaneuf-1978} ({\em empty circles}),
\protect\cite{Goffe-1979}
({\it empty squares}) and \protect\cite{s98} ({\it filled triangles}); c.e. cross section recommended
by \protect\cite{s98} ({\it solid line}).  {\it Bottom panel:} Collisional
rate coefficients: Langevin m.t. ({\it dashed line}), m.t. and c.e.({\it solid lines}).}
\label{cph}
\end{figure}

In the CNM and DiM, ion-neutral damping is dominated by the collisions
between neutral hydrogen and ionized carbon.  Fig.~\ref{cph} shows
the cross sections and reaction rates for m.t. and c.e. in the case
of collisions of C$^+$ ions with H atoms.  For collisions of C$^+$
ions with He atoms, no theoretical or experimental are available.
The m.t. transfer rate coefficient, according to the Langevin theory,
is $\langle\sigma_{\rm mt} v\rangle_{{\rm C}^+,{\rm He}}=1.33\times
10^{-9}$~cm$^3$~s$^{-1}$ (\citealt{pg08}).  
As in the case of H$^+$-- He
collisions, the large difference with the rate coefficients used
in this work resides in the assumption that the rate coefficient
is the same for all species.

\end{document}